\documentclass[12pt]{article}
\usepackage{amssymb}
\usepackage{amsmath}
\usepackage{amsthm}
\usepackage[round]{natbib}
\usepackage{graphicx}
\usepackage{eurosym}
\usepackage{amsfonts}
\usepackage{multirow}
\usepackage{paralist}
\usepackage{verbatim}
\usepackage{subfig}
\usepackage{mathrsfs}
\usepackage{bm}
\usepackage{xcolor}
\usepackage{hyperref}
\usepackage{tabularx}
\usepackage{booktabs} 
\usepackage{endnotes}
\usepackage[linesnumbered,lined,boxed,commentsnumbered]{algorithm2e}
\usepackage{soul}
\usepackage{enumitem} 
\usepackage{mathtools} 
\usepackage[normalem]{ulem}

\newtheorem{theorem}{Theorem} [section]

\newtheorem{definition}[theorem]{Definition}
\newtheorem{example}[theorem]{Example}

\newtheorem{proposition}[theorem]{Proposition}

\newtheorem{myprocedure}[theorem]{Procedure}
\newtheorem{casess}[theorem]{Case}
\allowdisplaybreaks[1]
\begin{document}

\author{Anjeza Bekolli\thanks{%
Department of Mathematics and Informatics, Faculty of Economy and
Agribusiness, Agricultural University of Tirana, Albania. e-mail:
abekolli@ubt.edu.al}, Luis A. Guardiola\thanks{%
Departamento de Matemáticas, Universidad de Alicante,
Spain. e-mail: luis.guardiola@ua.es} $^{,} $\thanks{%
Corresponding author.} \ and Ana Meca\thanks{I. U. Centro de Investigación Operativa, Universidad Miguel Hernández de Elche, 03202 Elche, Spain.
e-mail: ana.meca@umh.es}}
\title{HarvestTech agriculture cooperatives: Beneficiaries and compensations}
\maketitle

\begin{abstract}
%{\color{red} Agricultural industries face increasing pressure to optimize efficiency and reduce costs in a competitive and resource-constrained global market. As firms seek innovative ways to enhance productivity, cooperative strategies have emerged as a promising solution to address these challenges. In this context, game theory provides a powerful framework for analyzing and structuring such cooperative efforts, ensuring that each firm's contribution is fairly rewarded. This paper presents} an innovative approach to address challenges in agricultural crop processing through inter-firm cooperation. We introduce a new {\color{red}class of totally} balanced games that models strategic interactions among companies processing agricultural products. Our aim is to identify profit allocations that fairly compensate firms contributing to cost reduction and surplus processing for others. To achieve this, we will thoroughly examine the allocations resulting from each type of compensation and establish a compensation procedure that is coalitionarily stable. Our study demonstrates the feasibility and effectiveness of cooperative strategies for optimizing agricultural processes. Lastly, we will apply the findings to a {\color{red}case study}.

Agricultural industries face increasing pressure to optimize efficiency and reduce costs in a competitive and resource-constrained global market. As firms seek innovative ways to enhance productivity, cooperative strategies have emerged as a promising solution to address these challenges. In this context, game theory provides a powerful framework for analyzing and structuring such cooperative efforts, ensuring that each firm's contribution is fairly rewarded. This paper presents an innovative approach to address challenges in agricultural crop processing through inter-firm cooperation. A new class of totally balanced games is introduced, which models the strategic interactions among companies processing agricultural products. The objective is to identify profit allocations that fairly compensate firms contributing to cost reduction and surplus processing for others. To achieve this, the allocations resulting from each type of compensation will be thoroughly examined, and a coalitionally stable compensation procedure will be established. The study demonstrates the feasibility and effectiveness of cooperative strategies for optimizing agricultural processes. Lastly, the findings will be applied to a case study.


\end{abstract}

\newpage

\section{Introduction}
%Within our framework, a coalition enables each member to access the technologies owned
%by other members of the coalition. Consequently, the members can utilize the least costly
%technology available among the firms in the coalition.
%
%We assume that the different agriculture companies have specific quantities of product
%that they have already harvested. In other words, each company knows the exact amount
%in kilograms that it has available to process and convert into its final product. Moreover,
%this information (maximum harvest quantity, processing technology cost, and maximum processing
%capacity) is known by all members of the coalition.

Crop processing optimization plays an essential role in maximizing revenue and reducing losses in the changing agricultural industry.  This paper introduces an innovative framework that addresses the challenges faced in agricultural crop processing through inter-firm cooperation. The concept of inter-firm cooperation in agricultural processing is not entirely new. However, the application of cooperative game theory to model these interactions and optimize profit allocation is a novel approach. Cooperative game theory, particularly the concept of transferable utility (TU) games provides a robust framework for analyzing how coalitions of firms can share resources and optimize processes collectively. In a TU game, the value created by any coalition of players can be transferred among them, allowing for flexible profit-sharing arrangements that reflect each firm's contribution to the coalition's overall success. We propose a new type of balanced game that effectively represents the complexities of cooperation among firms processing agricultural products. This model aims to identify profit allocations that fairly compensate firms contributing to cost reduction and surplus processing for others. By examining various compensation scenarios, we establish a coalitionally stable compensation procedure, demonstrating the feasibility and effectiveness of cooperative strategies in optimizing agricultural processes. 

Our study is motivated by scenarios where agricultural firms face significant disparities between their harvesting and processing capacities. For instance, in Napa Valley, renowned worldwide for its wine production, companies like Napa Pure Juice Co., Golden Vineyards Inc., and Harvest Bliss Ltd. experience fluctuations in grape harvests and processing capacities. These companies produce similar products but operate with different capacities and cost structures, leading to inefficiencies and potential waste of resources. To illustrate the motivation behind our study, we consider in Section \ref{sec4} the hypothetical example of three companies in Napa Valley. Napa Pure Juice Co. owns its vineyards and also purchases grapes from small local producers.  

Our model considers a set of agricultural companies that harvest and process a single product. Each company knows the amount of harvested product it has available to process and convert into its final product. This information, including harvest quantity, processing technology cost, and maximum processing capacity, is known by all members of the coalition. This transparency is crucial for the validity of our cooperative model, as it ensures that all coalition members are fully informed and can make rational decisions based on complete information. The idea of our framework lies in the formation of coalitions among farmers, enabling them to access and utilize each other's processing technologies and capacities. This collaborative approach ensures that the least costly technology is available to all coalition members, thereby minimizing processing costs and maximizing overall profitability.  

In our cooperative game theory model, we define a HarvestTech (HT) situation, composed by a set of farmers that produce a single product in a single period of time, their processing capacities, harvested quantities, processing costs, and market price of the final product. In our model, farmers can cooperate in two primary ways: by sharing their processing technologies to utilize the least costly technology available within the coalition, and by sharing their processing capacities to ensure that all harvested quantities are processed efficiently. Formally, we define the cooperative game that arises from an HT situation and explore its properties, such as superadditivity and monotonic increasing characteristics, which ensure the viability of forming a grand coalition. 

These properties are crucial for the viability of forming a grand coalition and ensuring fair profit distribution. We propose the collaborative allocation (CO) method, which distributes profits based on each company's processing capacity or harvested quantity, depending on whether the total harvest can be fully processed. This method ensures that each company receives a fair share of the profits, maintaining stability within the coalition. However, the CO allocation may not adequately compensate companies that significantly improve the processing technology of others or those that assist in processing excess harvests. To address these issues, we introduce the “Best Technology Compensation” (BTC) and “Crop Reward Compensation” (CRC) allocations. The BTC allocation compensates companies with the lowest processing cost by taxing those with higher costs, while the CRC allocation compensates companies based on their processing capacity or harvested quantity. 

Our findings demonstrate that these cooperative strategies not only optimize agricultural processes but also ensure that all companies are fairly compensated for their contributions. The resulting allocations are coalitionally stable, promoting long-term cooperation among firms. 

%{\color{red}The paper is organized as follows. We begin with a lterature review and a preliminary
%section in cooperative game theory and other issues. In Section \ref{sec4},
%we introduce the crop and processing agriculture model. Then,
%Section 5 definimos el juego cooperativo asociado al modelo, y estudiamos sus principales propiedades. Además, proponemos una regla de reparto que is stable in the sense of the core. Sección 6 focuses on the study of a new allocation que compensen justamente aquellos agricultores que provocan una rebaja en los costes o aportan una capacidad de procesamiento mayor a la gran coalición. Section 7 presents a simulated case study of this cooperative model and the proposed allocation search procedure to, allowing us to compare different rules. Finally, Section 8 draws some
%conclusions and points out further research for scholars and
%practitioners in the field.}

The paper is organized as follows. It begins with a literature review and a preliminary section on cooperative game theory and related topics. In Section \ref{sec4}, the crop and agricultural processing model is introduced. Then, in Section 5, the cooperative game associated with the model is defined, and its main properties are analyzed. Additionally, a profit-allocation rule that is stable in the sense of the core is proposed. Section 6 focuses on a new allocation rules that fairly compensates farmers who contribute to cost reductions or provide greater processing capacity to the grand coalition. Section 7 presents a simulated case study of this cooperative model and the proposed allocation method, allowing for a comparison of different allocation rules. Finally, Section 8 draws conclusions and suggests avenues for further research for both scholars and practitioners in the field.

\section{Related literature}

In the literature, there are very few applications of the cooperative game theory model in the agricultural sector.  In this section we describe the literature related to our paper.

The study by \cite{tavanayi2021cooperative} introduces a cooperative cellular manufacturing system to reduce production costs by forming coalitions among companies, supported by a mathematical programming model and cooperative game theory methods. It is demonstrated by a case study of three high-tech suppliers to Mega Motor Company demonstrates significant cost savings and fair cost allocation using methods such as the Shapley value and the least core. Furthermore, the application of cooperative game theory extends beyond manufacturing, as exemplified by \cite{zheng2022cooperative} in the agricultural sector, where they develop an incentive mechanism through cooperative game theory to promote biopesticide adoption via Farmer Producer Organizations (FPOs), achieving superior economic and environmental outcomes through collective action.

Our paper investigates horizontal collaboration among agricultural companies using cooperative game theory to devise equitable methods for distributing gains among collaborating firms. Cooperative game theory has been extensively utilized to tackle various supply chain issues, as evidenced by numerous review papers focused on specific contexts. Examples include cooperative transportation \citep{cruijssen2007horizontal}, cooperative logistics network design \citep{hezarkhani2021collaboration}, cooperative inventory management \citep{fiestras2011cooperative,fiestras2012cost,fiestras2013new,fiestras2014centralized,fiestras2015cooperation}, cooperation in assembly systems \citep{bernstein2015cooperation}, cooperative sequencing \citep{curiel2002sequencing}, cooperative advertising \citep{jorgensen2014survey}, and cooperative distribution chains \citep{guardiola2007cooperation,GUARDIOLA2023102889} among others. Various surveys such as those by \cite{nagarajan2008game}, \cite{meca2008supply} and \cite{rzeczycki2022supply} provide insights into the applications of cooperative game theory in supply chain management.

Apple fruit cultivation is a significant agricultural activity in Albania, contributing substantially to the country's agrarian economy. The sector is characterized by small to medium-sized family-run orchards, which are predominant in the mountainous and hilly regions where the climate is conducive to apple growing. The main apple-producing regions include Korça, Dibër, and Shkodra. These areas benefit from favorable climatic conditions such as sufficient rainfall, adequate sunshine, and fertile soils, which are essential for high-quality apple production. 

The study by \cite{osmani1999efficiency} assesses technical efficiency and its influencing factors for apple farms in the Korca region using data from 150 randomly selected farmers and analyzed through the Stochastic Frontier Approach. This study emphasizes the need for improved extension services and policies supporting training, market access, and provision of quality inputs and technologies to enhance farmers’ skills and knowledge. The results also highlight that traditional orchards may be more efficient than modern ones due to the technical demands of the latter, suggesting a need for better technical support for modern orchard management. The study of \cite{canaj2024energy} is the first to examine the energy balance and environmental impacts of intensive apple production in Albania using Life Cycle Energy Analysis (LCEA) and Life Cycle Assessment (LCA). The study suggests that adopting precision agriculture, using organic fertilizers, investing in energy-efficient machinery, and providing training programs can enhance the sustainability of apple production in the region. This comprehensive assessment offers crucial insights into the energy dynamics and environmental repercussions of apple farming in Korça, providing a foundation for policy interventions aimed at promoting sustainable agricultural practices. Study by \cite{osmani2019small} evaluates the willingness to invest among small-scale apple farms in the Korca region of Albania and identifies key determinants. Using data from face-to-face interviews with apple farmers, descriptive statistics, and logistic regression analyses, the study finds that most farmers are inclined to invest due to positive investment climate expectations. Factors such as access to loans, advisory services, market competition, willingness to cooperate, and farm income positively influence investment willingness, while socio-demographic factors like age, education, and experience do not. Large farms show a greater propensity to invest. This assessment provides critical insights into the investment behavior of apple farmers in Korca, highlighting barriers and suggesting policy interventions to foster a better investment climate in the sector. 

\section{Preliminaries cooperative game theory}
In order to ensure clarity, we have included in this section the fundamental principles of cooperative game theory, which are essential for understanding and validating the findings presented in this paper.

A cooperative game with transferable utility (TU-game) comprises a set of players $N=\{1,2,...,n\}$ and a characteristic function $v$, which associates each subset of $N$ with a real number. The subsets of $N$ are referred to as coalitions, denoted by $S$. Formally, the characteristic function is a map $v:2^{N}\longrightarrow \mathbb{R}$ such that $v(\emptyset)=0$. The value $v(S)$ of the characteristic function represents the maximum profit that members of the coalition $S$ can achieve through cooperation. The coalition consisting of all agents, $N$, is defined as the grand coalition, where $\left\vert N\right\vert $ refers to the
cardinality of the grand coalition.  The subgame related to
coalition $S$, is the restriction of the map $v$ to the subcoalitions of $S$. 

A key question in cooperative game theory concerns how to fairly distribute the profits among members of the grand coalition once it has been formed. This distribution is accomplished through allocations, which are represented by a vector $x\in \mathbb{R}^{\left\vert N\right\vert }$. The class of superadditive games is especially intriguing, as it motivates the formation of the grand coalition by ensuring maximum profits for its members. Formally, the game is said to
be superadditive if the benefit obtained by the combination of any two
disjoint coalitions is at least as much as the sum of their individual
benefits. Specifically, $v(S\cup T)\geq v(S)+v(T)$ holds for all disjoint
coalitions $S,T\subseteq N$. Additionally, A TU-game $(N,v)$ is considered monotone increasing if larger coalitions
receive more significant benefits, which is expressed as $v(S)\leq v(T)$ for
all coalitions $S\subseteq T\subseteq N.$

The set of all vectors that efficiently allocate the profits of the grand
coalition and are coalitionally stable is referred to as the core of the
game $(N,v)$, which is denoted as $Core(N,v)$. More specifically, no player
in the grand coalition has an incentive to leave, and each coalition is
guaranteed to receive at least the profit allocated by the characteristic
function. Formally,
\begin{equation*}
Core(N,v)=\left\{ x\in \mathbb{R}^{\left\vert N\right\vert }:\sum_{i\in N}x_{i}=v(N)\text{
and }\sum_{i\in S}x_{i}\geq v(S)\ \text{\ for all }S\subset N\right\} .
\end{equation*}

In the core, when an element is proposed (hereafter referred to as a core allocation) where player
$i$ receives $x_{i}$, every coalition S of players should receive at least the amount they can generate on
their own. An allocation $x$ can also be described as stable in the sense of the core.

A TU-game is classified as balanced only when the core is nonempty,
as detailed in \cite{bondareva1963some} and \cite{shapley1967balanced}. If the core of every subgame is
nonempty, the game $(N,v)$ is considered to be a totally balanced game (see \cite%
{shapley1969market}). A game $(N,v)$ is regarded as convex if for all $i\in N$ and all $%
S,T\subseteq N$ such that $S\subseteq T\subset N$ with $i\in S,$ then $%
v(S)-v(S\setminus \{i\})\geq v(T)-v(T\setminus \{i\}).$ It is widely
acknowledged that convex games are superadditive, and superadditive games
are totally balanced. Shapley establishes in \cite{shapley1971cores} that the core of
convex games is large enough.

\section{Crop and processing agriculture model} \label{sec4}

In Napa Valley, renowned worldwide for its wine production, there are
several companies dedicated to processing grapes into grape juice and wine. We will focus on three fictional companies, based on typical operations in
the region: Napa Pure Juice Co., Golden Vineyards Inc., and Harvest Bliss
Ltd. We consider a common product for the three companies to be table wine. These wines are generally more affordable and are sold within a similar price range.

\begin{enumerate}
\item Napa Pure Juice Co.: This company owns its vineyards and also
purchases grapes from small local producers. Its primary objective is to
produce pure, additive-free table wine. It has a processing capacity of
253,000 kilograms of grapes per season, but in the last harvest, the
quantity of grapes obtained was 198,000 kilograms.

\item Golden Vineyards Inc.: This company primarily produces artisanal table
wines. It has a smaller processing capacity of 259,000 kilograms per season
and harvested 189,000 kilograms of grapes this year. Like Napa Pure Juice
Co., Golden Vineyards Inc. also purchases grapes from local producers to
supplement its production.

\item Harvest Bliss Ltd.: This company focuses on producing organic table wine. Its processing capacity is 137,000 kilograms, and its recent
harvest was 229,000 kilograms. Harvest Bliss Ltd. prioritizes purchasing
grapes from certified organic farms to ensure the quality of its product.
\end{enumerate}

During the last harvest season, all of these companies faced a common
dilemma: the quantity they harvested is different from the quantity they can
process. To maximize their profits and avoid wasting grapes, they decided
to explore cooperation.

\begin{figure}
\centering
\includegraphics[width=9cm]{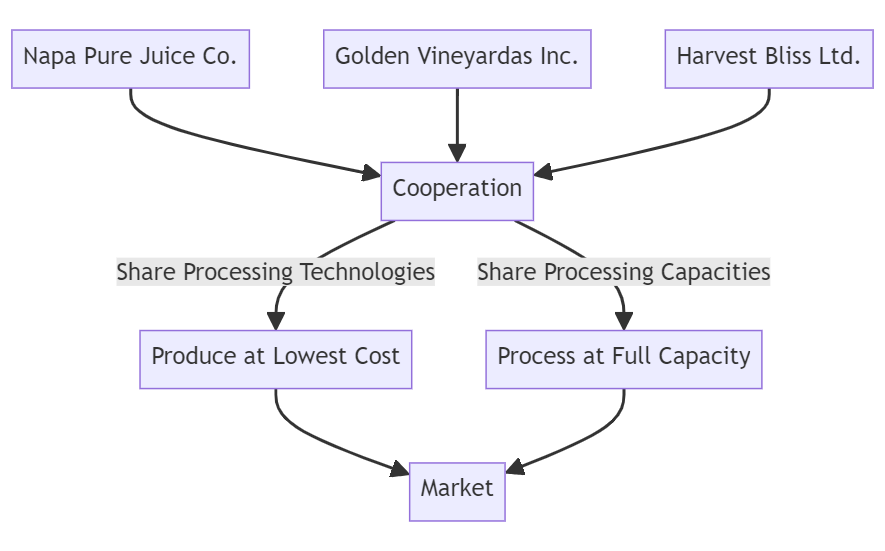}
\caption{Cooperation scheme}
\label{fig:esq}
\end{figure}

The companies suggested sharing processing capacity and technology.
Napa Pure Juice Co., with an excess capacity of 55,000 kilograms of grapes,
decided to share this capacity with Harvest Bliss Ltd., which had a deficit
of 92,000 kilograms, and with Golden Vineyards Inc., which had a deficit of
70,000 kilograms. Furthermore, Napa Pure Juice Co. had a processing cost of
\$0.8 per kilogram, Golden Vineyards Inc. of \$0.95, and Harvest Bliss Ltd. of
\$0.9. The market price of grape juice and wine remained at \$1.70 per
kilogram. This collaboration is not a trivial task, as it requires the
cooperation of all companies in terms of their processing capacities and
technologies, as well as the precise evaluation of each company's
contribution to ensure fair compensation. We consider a cooperative game
theory model that helps these companies resolve their dilemma. In Figure \ref{fig:esq} we present an outline of the cooperation between the companies.

%\begin{center}
%\includegraphics[width=9cm, height=6cm]{market.png} 
%\end{center}

In this paper, we focus on agricultural companies that harvest and process a
single product. These companies may obtain their harvest from their own
plantations or by purchasing from various farmers. However, we focus only on the cost of processing the harvest and the quantity
processed. Real-life examples of such companies could be juice processing
companies or those that clean, sort, and package fruits and vegetables. In
our model, we assume that there is a set of $N=\{1,...,n\}$ companies
that obtain the same product. We assume that there is a single harvesting
period during which each company $i\in N$ harvests a quantity $Q_{i}$ in
kilograms and also has a maximum processing capacity $K_{i}$ during that
period, which is also measured in kilograms. We assume that there exists at
least one company $i\in N$ such that $K_{i}\neq Q_{i}$ (in the case where $K_{i}=Q_{i}$ for all players, there is no incentive for them to cooperate with each other). The final product, whether in liters or
packages, can be sold in the market at a price $p>0$ per kilogram and the
individual unit processing costs are $c_{i}<p$ for all $i\in N$. Without
loss of generality, we assume that the processed kilograms are sold
without wasting any part of the harvest during processing, as we can
directly deduct that percentage from $Q_{i}${} if necessary. The profit
obtained by an individual company $i\in N$ can be calculated as follows: $%
(p-c_{i})\cdot \min \{K_{i},Q_{i}\}$, where $(p-c_{i})$ is the profit
obtained from the sale of each kilogram of product, multiplied by the number
of kilograms processed, which will always be the minimum between the
quantity of the harvest and the number of kilograms that can be processed.

In this context, companies cooperate in two ways:

\begin{enumerate}
\item Sharing their processing technologies, so that if we consider a
coalition of companies $S\subseteq N$, they all produce at the lowest cost $%
c_{S}$ among all the members of the coalition; i.e., $c_{S}:=\underset{i\in S%
}{\min }\{c_{i}\}.$

\item Companies in the same coalition with surplus processing capacity can help process the harvests of those whose harvest quantity exceeds their own processing capabilities by sharing their resources. That is to say, $\min
\{K_{S},Q_{S}\}$ where $K_{S}:=\sum_{i\in S}K_{i}$ and $Q_{S}:=\sum_{i\in
S}K_{i}.$
%By sharing their processing capacity, companies within the same
%coalition that have more processing capacity than harvest quantity can
%process the harvest of other companies in the coalition whose harvest
%quantity exceeds their processing capacity. 
\end{enumerate}

Formally, we define a HarvestTech situation as a tuple $(N,K,Q,C,p)$
(hereinafter referred to as an HT situation), where $K=(K_{1},...,K_{n})$
represents the maximum processed capacity in kilograms for each player, $%
Q=(Q_{1},...,Q_{n})$ represents the maximum harvested quantity (in
kilograms) for each player, $C=(c_{1},...,c_{n})$ represents the total cost
of processing each kilogram of harvest for each player. Table \ref{tabla:datos} provides a summary of the processing quotas and limitations and quantity harvested for each of the Napa Valley companies.

\begin{table}[h!]
\centering
\begin{tabular}{|c|c|c|c|}
\hline
Companies & \(C\) & \(K\) & \(Q\) \\ \hline
Napa Pure Juice Co. & 0.95 \$ & 253000 Kg. & 198000 Kg. \\ \hline
Golden Vineyards Inc. & 0.80 \$ & 259000 Kg. & 189000 Kg. \\ \hline
Harvest Bliss Ltd. & 0.90 \$ & 137000 Kg. & 229000 Kg. \\ \hline
\end{tabular}
\caption{Production cost, processing and harvest for Napa Valley}
%\caption{Production costs, maximum amount of processing and harvest obtained}
\label{tabla:datos}
\end{table}

Once the model has been described, we turn to an analysis of coalition formation among farmers that enables them to access and utilize each other’s processing technologies and capacities.

\section{HarvestTech cooperative game}

We aim to study the cooperation model presented in the previous section by using cooperative game theory. We define the cooperative game that arises from an HT situation as follows.

\begin{definition}
Let $(N,K,Q,C,p)$ be an HT situation. The corresponding game $(N,v)$ with
characteristic function $v:\mathcal{P}(N)\rightarrow \mathbb{R}_{+}$ is
given by 
\begin{equation*}
v(S):=(p-c_{S})\cdot \min \{K_{S},Q_{S}\}
\end{equation*}%
for all coalitions $S\subseteq N$, $v(\emptyset )=0.$ 
\end{definition}

From now on the game $%
(N,v)$ is called HT game. With the aim of understanding how these games are generated, we calculate the game associated with the cooperation situation of the three companies that cultivate grapes.

\begin{example}
\label{ejemplo 1} Consider a HT situation with the above three  companies, where: $%
K=(253000,259000,137000)$, $Q=(198000,189000,229000)$, $C=(0.95,0.8,0.9)$ and $p=1.70.$ Table \ref{tabla:game} shows as the
cooperative HT game associated with this situation.

\begin{table}[h!]
\centering
\begin{tabular}{|c|c|c|c|c||c||}
\hline
& $c_{S}$ & $p-c_{S}$ & $K_{S}$ & $Q_{S}$ & $v(S)$ \\ \hline
$\{1\}$ & 0.95 & 0.77 & 253000 & 198000 & 148500 \\ \hline
$\{2\}$ & 0.8 & 0.9 & 259000 & 189000 & 170100 \\ \hline
$\{3\}$ & 0.9 & 0.8 & 137000 & 229000 & 109600 \\ \hline
$\{1,2\}$ & 0.8 & 0.9 & 512000 & 387000 & 348300 \\ \hline
$\{1,3\}$ & 0.9 & 0.8 & 390000 & 427000 & 312000 \\ \hline
$\{2,3\}$ & 0.8 & 0.9 & 396000 & 418000 & 356400 \\ \hline
$\{1,2,3\}$ & 0.8 & 0.9 & 649000 & 616000 & 554400 \\ \hline
\end{tabular}
\caption{HT game for Napa Valley}
\label{tabla:game}
\end{table}

As we can see in this example, company 2 (Golden Vineyards Inc.) possesses the cheapest processing technologies and can process its entire harvest. However, company 3 (Harvest Bliss Ltd.) has a higher processing cost and does not have the capacity to process its entire harvest. Meanwhile, company 1 (Napa Pure Juice Co.) has the highest processing cost but can process its entire harvest. By joining forces, they manage to process the entire joint harvest at a minimal cost. Consequently, the profit generated by the grand coalition surpasses that generated in any sub-coalition and even that of the sum of any partition of coalitions (superadditivity). Note that the allocation $(178,200;170,100;206,100)$ is a core allocation, and so the HT game is balanced, but no convex (i.e., $198,000=v(\{1,2,3\})-v(\{2,3\})<v(\{1,3\})-v(\{3\})=202,400$).

%As we can see in this example, player 2 (Golden Vineyards Inc.) possesses {\color{red} the cheapest} processing technologies
%but has a lowest harvest, which
%prevents them from processing their entire harvest. However, the opposite
%situation occurs for the rest of the players. By joining forces, they manage
%to process almost the entire joint harvest at a minimal cost. Consequently,
%the profit generated by the grand coalition surpasses that generated in any
%sub-coalition and even that of the sum of any disjoint union of coalitions
%(superadditivity). 
\end{example}

The following proposition demonstrates the main properties of this class of
games.

\begin{proposition}
\label{prop1}
Let $(N,K,Q,C,p)$ be an HT situation and $(N,v)$ be the corresponding HT game $%
(N,v)$. Then,

\begin{itemize}
\item[(i)] $v(S)\geq 0$ for all $\emptyset \neq S\subseteq N.$

\item[(ii)] $v$ is superadditive;

\item[(iii)] $v$ is monotone increasing.
\end{itemize}
\end{proposition}

The Appendix provides the proofs of all results. Property (i) ensures that no company incurs losses. Properties (ii) and (iii) are crucial for the viability of forming a grand coalition. It is noteworthy that in superadditive games,
it is reasonable for the grand coalition to form. This is because the
benefit acquired by the grand coalition is at least as great as the sum of
the benefits of any other coalition and its complement, i.e., $v(N)\geq
v(S)+v(N\setminus S),$ for all $S\subseteq N.$
Our next objective will be to examine whether HT games consistently have coalitionally stable profit allocations. Inspired by the previous example, we define the following allocation:

\begin{definition}
Let $(N,K,Q,C,p)$ be an HT situation and $(N,v)$ be the corresponding HT game $%
(N,v)$. We define the collaborative allocation (CO allocation) as $\gamma (v)=\left( \gamma
_{i}(v)\right) _{i\in N}$ where,
\begin{equation*}
\gamma _{i}(v)=\left\{ 
\begin{array}{cc}
(p-c_{N})\cdot K_{i} & if\text{ }K_{N}\leq Q_{N}, \\ 
(p-c_{N})\cdot Q_{i} & if\text{ }K_{N}>Q_{N}.%
\end{array}%
\text{ }\right.
\end{equation*}
\end{definition}
\bigskip
The CO allocation proposes to distribute the profit generated by the grand coalition as follows: if the total harvest cannot be fully processed ($K_{N}\leq Q_{N}$), then the profit assigned to each company is the profit obtained from selling the maximum quantity it can individually process at the lowest processing cost among all companies. Conversely, if the total harvest of the grand coalition can be fully processed ($K_{N}>Q_{N}$), then each company is allocated its harvested quantity at the lowest possible processing cost.

The next theorem shows that the CO allocation is coalitionally stable in the sense of the core.
\begin{theorem}
\label{prop2}
Let $(N,K,Q,C,p)$ be an HT situation and $(N,v)$ be the corresponding HT game $%
(N,v)$. Then, $\gamma (v)\in Core(N,v).$
\end{theorem}

Following the above theorem, HT games are balanced. It is easy to see that every subgame of
HT games is also an HT game. Thus, HT games are totally balanced.
%\begin{corollary}
%HT games are totally balanced.
%\end{corollary}

The reader may notice that for the three companies in Example \ref{ejemplo 1}, the CO allocation results in $(178,200;170,100;206,100)$. However, despite being a core allocation, it does not adequately compensate Golden Vineyards Inc. (Player 2), which shares its processing technology, the cheapest among the three companies, with the others. Additionally, under the CO allocation, each company profits based on the amount of harvest they have individually obtained, but this is not entirely fair. Harvest Bliss Ltd. (Player 3) relies on the processing capacities of the other two companies to convert its harvest into a marketable final product, and therefore, it should also compensate the other companies.

In the following section, we introduce two new allocations that compensate the players by addressing the two unfair behaviors described above. Finally, we will provide a procedure to combine these two compensations.

\section{Beneficiaries and compensations}

The CO allocation does not adequately compensate players who improve the processing technology of others, those who assist in processing others' harvests (in case of $K_{N}> Q_{N}$) or those with larger harvests (in case of $K_{N}\leq Q_{N}$). Therefore, one might doubt the willingness of these players to join and share their technologies, processing capacities, and harvests. Below, we present two modifications to this CO allocation to remedy this issue. These allocations charge an amount proportional to the gain received by players who benefit the most from the assistance of others and redistribute it among those who have a positive impact on the profits of others.

The first modification to the CO allocation is to tax the companies that do not have the lowest processing cost (with a fixed percentage rate) and distribute the collected amount among the company or group of companies that possesses the lowest possible processing cost. In this way, we compensate those companies that increase the profits of others by allowing them to operate with the best technology.
\begin{definition}
Let $(N,K,Q,C,p)$ be an HT situation and $(N,v)$ be the corresponding HT game $%
(N,v)$. Given $\alpha \geq 0$, 
the Best Technology Compensation allocation (henceforth BTC allocation) $%
T(v,\alpha )=\left( T_{i}(v,\alpha )\right) _{i\in N}$ is defined as follows:

\begin{equation*}
T_{i}(v,\alpha ):=\left\{ 
\begin{array}{ll}
\gamma _{i}(v)-\alpha \cdot \gamma _{i}(v), & i\in N\backslash M, \\ 
&  \\ 
\gamma _{i}(v)+\frac{\alpha }{\left\vert M\right\vert }\sum_{j\in N/M}\gamma
_{j}(v), & i\in M,%
\end{array}%
\right.
\end{equation*}

\noindent where $M:=\{i\in N:i\in \underset{j\in N}{\arg \min }\{c_{j}\}\}$.
\end{definition}

The following result provides a necessary and sufficient conditions for the BTC allocation to belong to the core of the game.

\begin{theorem}
\label{theo1}Let $(N,K,Q,C,p)$ be an HT situation and $(N,v)$ be the
corresponding HT game $(N,v)$. Then, $T(v,\alpha )$ belongs to the core if and
only if $\ $%
\begin{equation*}
0\leq \alpha \leq \min \left\{ \underset{S\subseteq N\backslash M}{\min }%
\left\{ 1-\frac{v(S)}{\underset{i\in S}\sum\gamma _{i}(v)}\right\} ,\underset{S\in
\Lambda }{\min }\left\{ \frac{\underset{i\in S}\sum\gamma _{i}(v)-v(S)}{\underset{i\in
S\cap (N\backslash M)}\sum\gamma _{i}(v)-\frac{\left\vert S\cap M\right\vert }{%
\left\vert M\right\vert }\underset{j\in N/M}\sum\gamma _{j}(v)}\right\} \right\},
\end{equation*}%
where $\Lambda :=\left\{ S\subseteq N:S\cap M\neq \emptyset ,S\cap
(N\backslash M)\neq \emptyset \text{ and}\underset{i\in
S\cap (N\backslash M)}\sum\gamma _{i}(v)>\frac{\left\vert S\cap M\right\vert }{\left\vert
M\right\vert }\underset{j\in N/M}\sum\gamma _{j}(v)\right\}. $
\end{theorem}
\medskip

The second modification to the CO allocation divides the companies into two sets: those whose processing capacity exceeds the harvest amount, and those whose harvest amount exceeds their processing capacity. Depending on the allocation criterion of the CO allocation (whether it allocates based on processing capacity or harvested quantity), this rule compensates one set of players or the other.

\begin{definition}
Let $(N,K,Q,C,p)$ be an HT situation and $(N,v)$ be the corresponding HT game $%
(N,v)$. Consider $E:=\{i\in N:Q_{i}=K_{i}\}\neq N$ and 
\begin{equation*}
H:=\left\{ 
\begin{array}{ll}
\{i\in N:K_{i}<Q_{i}\} & \text{if }K_{N}\leq Q_{N}, \\ 
&  \\ 
\{i\in N:K_{i}>Q_{i}\}, & \text{if }K_{N}>Q_{N}.%
\end{array}%
\right.
\end{equation*}%
Given $\beta \geq 0$, we define the Crop Reward Compensation allocation (henceforth CRC allocation) $%
R(v,\beta )=\left( R_{i}(v,\beta )\right) _{i\in N}$ as:
\begin{equation*}
R_{i}(v,\beta ):=\left\{ 
\begin{array}{ll}
\gamma _{i}(v)-\beta \cdot \gamma _{i}(v), & i\in N\backslash \left( H\cup
E\right) , \\ 
\gamma _{i}(v) & i\in E, \\ 
\gamma _{i}(v)+\frac{Q_{i}-K_{i}}{Q_{H}-K_{H}}\beta \sum_{j\in N\backslash
H}\gamma _{j}(v), & i\in H.%
\end{array}%
\right.
\end{equation*}%
\end{definition}
\bigskip
The reader may notice that CRC allocation is well-defined because of $%
Q_{H}-K_{H}\neq 0$. Moreover, $H\neq \emptyset$ since if $K_{N}\leq Q_{N}$ there must exist at least one player $i \in N$ who satisfies $K_{i}\leq Q_{i}$ ($E\neq N$) and the same occurs for case $K_{N}>Q_{N}$.
To understand this second modification, we must consider the two possible scenarios that may arise. In case $K_{N}\leq Q_{N}$, the CO allocation has distributed the profits according to each player's processing capacity, and the set $H$ would be formed by players with less production capacity than harvest capacity. These players would process their surplus harvest with those players who satisfy $Q_{i} < K_{i}$ (the set $N\backslash \left( H\cup E\right)$). Similarly, in case $K_{N}>Q_{N}$, the set $H$ would now consist of players with surplus processing capacity. Since in this case the CO allocation has distributed profits based on the harvested quantity, it is the players who satisfy $Q_{i} > K_{i}$ who should compensate them for processing part of their harvest. 
The following result provides the necessary and sufficient condition for the CRC allocation to belong to the core of the game.

\begin{theorem}
\label{theo2}Let $(N,K,Q,C,p)$ be an HT situation and $(N,v)$ be the
corresponding HT game $(N,v)$. Then, $R(v,\beta )$ belongs to the core if and
only if%

\[
\resizebox{1\textwidth}{!}{$
0\leq \beta \leq \min \left\{ \underset{S\subseteq N\backslash H:S\backslash
E\neq \emptyset }{\min }\left\{ 1-\frac{v(S)}{\underset{i\in S\backslash
E}\sum\gamma _{i}(v)}\right\} ,\underset{S\in \Pi }{\min }\left\{ \frac{%
\underset{i\in S}\sum\gamma _{i}(v)-v(S)}{\underset{i\in S\cap \left( N\backslash \left(
H\cup E\right) \right)}\sum\gamma _{i}(v)-\frac{Q_{S\cap H}-K_{S\cap H}}{%
Q_{H}-K_{H}}\underset{j\in N\backslash \left( H\cup E\right)}\sum\gamma _{j}(v)}%
\right\} \right\}
$}
\]
where, 
\small{
\begin{equation*}
\Pi :=\left\{ S\subseteq N:%
\begin{array}{c}
S\cap \left( N\backslash \left( H\cup E\right) \right) \neq \emptyset ,S\cap
H\neq \emptyset \text{ and} \\ 
\sum_{i\in S\cap \left( N\backslash \left( H\cup E\right) \right) }\gamma
_{i}(v)>\frac{Q_{S\cap H}-K_{S\cap H}}{Q_{H}-K_{H}}\sum_{j\in N\backslash
\left( H\cup E\right) }\gamma _{j}(v)%
\end{array}%
\right\} 
\end{equation*}}
\end{theorem}
\medskip

We apply these two new allocations concepts to the three companies in Napa Valley. First, we need to determine the thresholds $\alpha$ and $\beta$. To do this, we apply Theorems \ref{theo1} and \ref{theo2} and find that $0\leq \alpha \leq 0.1666$ and $0\leq \beta \leq 0.2183$. This means that the maximum compensation percentage for the BTC allocation should be at most approximately $16.66\%$, and for the CRC allocation, approximately $21.83\%$. We take half of these values: 8.33\% and 10.91\%. In Table \ref{tabla:comp}, we can compare the CO allocation with its two variations:

\begin{table}[h!]
\centering
\begin{tabular}{|c|c|c|c|}
\hline
Companies & $\gamma (v)$ & $T(v,0.0833)$ & $R(v,0.1091)$ \\ \hline
Napa Pure Juice Co. & 178200  & 163350 & 188100 \\ \hline
Golden Vineyards Inc. & 170100 & 202125 & 182700 \\ \hline
Harvest Bliss Ltd. & 206100  & 188925 & 183600 \\ \hline
\end{tabular}
\caption{Comparison of the CO, BTC, and CRC allocations}
\label{tabla:comp}
\end{table}

However, while each variation of the CO allocation compensates players who either contribute to reducing costs for others or processes surplus for others, they do not perform both of these compensations simultaneously. Therefore, our next objective is to attempt to combine them. To achieve this, we will carefully review Table \ref{tabla:sen}, which illustrates the direction of compensations made by the BTC and CRC allocations.

\begin{table}[h!]
\centering
\begin{tabular}{|c|c|c||c|}
\hline
Companies & \tiny{$T(v,0.0833)-\gamma (v)$} & \tiny{$R(v,0.1091)-\gamma (v)$} & \tiny{$T(v,0.0833)+R(v,0.1091)-2\gamma (v)$} \\ \hline
Napa Pure Juice Co. & -14850  & +9900 & -4950 \\ \hline
Golden Vineyards Inc. & +32025 & +12600 & 44625 \\ \hline
Harvest Bliss Ltd. & -17175  & -22500 & -39675\\ \hline
\end{tabular}
\caption{Net compensations among players for BTC, CRC allocations}
\label{tabla:sen}
\end{table}

As we can observe, both Napa Pure Juice Co. and Harvest Bliss Ltd. must compensate Golden Vineyards Inc. with approximately $8.33\%$ of their profits for sharing with them the best technology and reducing their processing costs. Additionally, Harvest Bliss Ltd. must compensate the other two companies with approximately $10.91\%$ of its profits for processing their surplus harvests. Note that the distribution is proportional to each company's surplus processing capacity.

An intuitive idea, in seeking a distribution that incorporates both compensations, is to sum the net compensations, as we have done in the third column of Table \ref{tabla:sen}. If we apply these final compensations to the CO allocation, we obtain the distribution $(173,250;214,725;166,425)$, which proves to be coalitionally stable.
\medskip

The following procedure describes how to construct an allocation, taking into account all the types of compensations that we have considered previously.

%\begin{definition}
%Let $(N,K,Q,C,p)$ be an HT situation and $(N,v)$ be the corresponding HT game $%
%(N,v)$. We define the HarvestTech Reward allocation (HTR allocation) as $%
%HT(v,\alpha ,\beta ):=T(v,\alpha )+R(v,\beta )-\gamma (v).$
%\end{definition}

\begin{myprocedure}\label{pro}
Let $(N,K,Q,C,p)$ be an HT situation and $(N,v)$ be the corresponding HT game $(N,v)$. Next, we will detail the procedure to obtain this allocation: 
\begin{enumerate}
%\item Calculate $c_{N}$ and $\min \{K_{N},Q_{N}\}$.
\item Calculate $\gamma (v)$.
%\item Obtain $M=\{i\in N:i\in \underset{j\in N}{\arg \min }\{c_{j}\}\}$.
\item Determine $\bar{\alpha}$, the threshold of $\alpha$, by using Theorem \ref{theo1}.
%\item Calculate $T(v,\alpha ^{*})$ with $\alpha ^{*} \leq \frac{\bar{\alpha}}{2}$.
\item Determine $\bar{\beta}$, the threshold of $\beta$, by using Theorem \ref{theo2}.
%\item Calculate $R(v,\beta^{*})$ with $\beta^{*} \leq \frac{\bar{\beta}}{2}$.
%\item Calculate $HTR(v,\alpha^{*}  ,\beta^{*}  ):=T(v,\alpha^{*}  )+R(v,\beta^{*} )-\gamma (v)$.
\item Find an allocation depending on the values of $\bar{\alpha}$ and $\bar{\beta}$:
\begin{itemize}
    \item If \(\bar{\alpha} = 0\) and \(\bar{\beta} = 0\), the best allocation is  $\gamma(v)$.
    \item If \(\bar{\alpha} > 0\) and \(\bar{\beta} = 0\), the best allocation is $T(v,\alpha ^{*})$ with $\alpha ^{*} \leq \frac{\bar{\alpha}}{2}$.
    \item If \(\bar{\alpha} = 0\) and \(\bar{\beta} > 0\), the best allocation is $R(v,\beta^{*})$ with $\beta^{*} \leq \frac{\bar{\beta}}{2}$.
    \item If \(\bar{\alpha} > 0\) and \(\bar{\beta} > 0\), the best allocation is: $$HTR(v,\alpha^{*}  ,\beta^{*}  ):=T(v,\alpha^{*}  )+R(v,\beta^{*} )-\gamma (v).$$ with $\alpha ^{*} \leq \frac{\bar{\alpha}}{2}$ and $\beta^{*} \leq \frac{\bar{\beta}}{2}$.
\end{itemize}

\end{enumerate}
\end{myprocedure}

From now on, we refer to the HarvestTech Reward allocation as the HTR allocation, which is the resulting allocation from this procedure. The following result shows that the HTR allocation always belongs to the core of the HT game.

\begin{theorem}
\label{theo3}Let $(N,K,Q,C,p)$ be an HT situation and $(N,v)$ be the
corresponding HT game $(N,v)$.  Then, HTR allocation obtained with Procedure \ref{pro} is a core allocation
%If $ \alpha ^{*} \leq \frac{\bar{\alpha}}{2}$ and $\beta^{*} \leq \frac{\bar{\beta}}{2}$ then, $HTR(v,\alpha ^{*},\beta^{*} )$ belongs
%to the core.
\end{theorem}

This allocation builds upon the CO allocation and incorporates the two compensations seen earlier, BTC and CRC. Therefore, all companies participating in the cooperation will be compensated for their contributions to other players, both in cost reduction and in processing the surpluses of those players unable to process their entire harvest.
\medskip

In the following section, we provide a case study of apple farming in Albania, for which we apply the procedure described earlier with the aim of finding a stable allocation in the sense of the core.

%and the proposed allocation search procedure to a {\color{red}\sout{real-world}} context.

\section{\protect\bigskip Case studies of apple farming in Albania}

%Apple fruit cultivation is a significant agricultural activity in Albania, contributing substantially to the country's agrarian economy. The sector is characterized by small to medium-sized family-run orchards, which are predominant in the mountainous and hilly regions where the climate is conducive to apple growing. The main apple-producing regions include Korça, Dibër, and Shkodra. These areas benefit from favorable climatic conditions such as sufficient rainfall, adequate sunshine, and fertile soils, which are essential for high-quality apple production. 

Apple farming is an essential source of income for many rural households in Albania. The sector not only supports local economies through direct sales but also generates employment opportunities in related activities such as harvesting, packaging, and distribution. The domestic market is the primary consumer of Albanian apples, although there is potential for expanding exports to neighboring countries. 

Our case study is concentrated in the Korça region, where a wide range of apple varieties are cultivated, including Golden Delicious, Starking, Granny Smith, Idared, and Fuji. Farmers in the region use modern agricultural practices such as integrated pest management (IPM), drip irrigation, and proper pruning techniques to maximize yield and improve fruit quality. 

Approximately 3,500 hectares of land in Korça are dedicated to apple orchards, and the region produces around 80,000 to 90,000 tons of apples annually, which accounts for about 65-70\% of Albania's total apple production. The average yield in Korça is about 25–30 tons per hectare, which is higher than the national average due to the region's favorable growing conditions and advanced farming techniques. 

Refrigerated warehouses play a crucial role in the apple fruit supply chain in the Korça region. They help preserve the quality and extend the shelf life of apples, allowing farmers to store their produce and sell it throughout the year rather than just during the harvest season. Many apple farmers in Korça have invested in refrigerated storage facilities to enhance their storage capabilities and market their products more effectively. There are approximately 50 to 60 refrigerated warehouses in the Korça region, primarily operated by apple farmers or farmer cooperatives. 

The total storage capacity of these refrigerated warehouses in Korça is estimated to be around 40,000 to 50,000 tons of apples. Individual warehouses vary in capacity, with smaller facilities storing around 200–500 tons and larger ones up to 1,000 tons or more. Most refrigerated warehouses in Korça are equipped with modern cooling technology, including controlled atmosphere storage systems that regulate oxygen and carbon dioxide levels to slow down the ripening process. Advanced temperature and humidity control systems ensure optimal storage conditions, maintaining temperatures between 0 and 4 degrees Celsius and relative humidity levels around 90–95\%.

The availability of refrigerated storage has significantly boosted the income of apple farmers in Korça by reducing post-harvest losses and allowing for better price negotiation. 

It has also facilitated the export of apples to international markets by ensuring the quality of the fruit during transit. Approximately 80–90\% of apple farmers in Korça utilize refrigerated warehouses for storing their produce. During peak harvest season, these facilities operate at near full capacity, while in off-peak seasons, they store apples to meet market demand and supply contracts. There has been a noticeable increase in investment in refrigerated storage infrastructure over the past decade, supported by both private investments and government subsidies. International development programs and grants have also contributed to the modernization and expansion of storage facilities in the region (see \cite{shkreli2019citrus}).

Establishing and operating refrigerated warehouses for apple fruit storage involves several costs, including initial investment, operational expenses, and maintenance. The initial investment covers the cost of constructing and installing modern refrigeration systems, including compressors, condensers, evaporators, and controlled atmosphere systems. Another important aspect is proper insulation and sealing to ensure energy efficiency, along with the cost of obtaining necessary permits and adhering to regulatory requirements. 
Operational expenses include electricity, which is a significant operational cost due to the energy-intensive nature of refrigeration, and hire skilled personnel to manage and operate the warehouse, including maintenance staff and operators.

In addition, insurance coverage for the facility, equipment, and stored produce. For some farmers, these costs are reduced because they have invested in energy-efficient equipment and renewable energy sources, such as solar panels. Some farmers have benefited from leveraging available government programs and international grants, and they also provide training for staff on best practices and implementing automation. It is also important to note that some of them have applied cooperative models while sharing warehouse facilities and costs among multiple farmers, resulting in lower individual expenses and improved utilization rates. 

The cost per kilogram varies from one farmer to another. This is because smaller warehouses may have higher per kilogram costs due to less efficient use of space and resources. Farmers who invest in energy-efficient refrigeration systems and renewable energy sources (e.g., solar panels) can significantly lower their electricity costs. Those using outdated or less efficient equipment may incur higher energy costs, raising the overall storage cost per kilogram. Warehouses that use automated systems and technology to reduce labor requirements have lower operational costs. Farmers who do not invest in regular maintenance face higher costs due to breakdowns and inefficiencies. Independent farmers managing their own facilities without sharing face higher per-kilogram costs due to lower utilization rates. Farmers who receive government subsidies or grants for building and operating refrigerated warehouses reduce their overall costs. Warehouses located closer to markets or distribution centers might incur lower transportation and logistics costs. 

Table \ref{tab:data22} shows a database processed by the Regional Agency of Agricultural Extension, Korca. In the official information requested by the authors, a simple database was made available with the names of farmers who produce and store apple fruit only in the district of Korce. In the database, we have information about the municipality and the village to which the farmer belongs, together with their contact number. 

To maintain the privacy of the owners, in Table \ref{tab:data22} we have replaced the names of the companies with numbers at the request of the Regional Agency of Agricultural Extension, Korca.

We must emphasize the fact that we are dealing with a product that has variability in the amount of production from one season to another, including fluctuations in the selling price or the calculation of costs. Our goal is to provide an application model that generates results for every input that we can provide.

\begin{table}[h]
    \centering
    \begin{minipage}{0.48\textwidth}
        \centering
        \resizebox{\textwidth}{!} {%
        \begin{tabular}{cccc}
            \toprule
            \small{\textbf{Agri-Owner}} & \small{\textbf{Harvest Quantity}} &  \small{\textbf{Processing Capacity}} & \small{\textbf{Processing Cost}} \\
            \midrule
            1  & 210000  & 400000  & 0.46909 \\
            2  & 145000  & 300000  & 0.49732 \\
            3  & 96000   & 300000  & 0.45858 \\
            4  & 70000   & 420000  & 0.46159 \\
            5  & 63000   & 120000  & 0.49479 \\
            6  & 90000   & 120000  & 0.54098 \\
            7  & 80000   & 60000   & 0.47191 \\
            8  & 73000   & 140000  & 0.50614 \\
            9  & 60000   & 70000   & 0.48823 \\
            10 & 100000  & 87000   & 0.52852 \\
            11 & 120000  & 70000   & 0.46233 \\
            12 & 68000   & 110000  & 0.49598 \\
            13 & 75000   & 500000  & 0.51158 \\
            14 & 120000  & 155000  & 0.52802 \\
            15 & 72000   & 200000  & 0.54133 \\
            16 & 66000   & 240000  & 0.49758 \\
            17 & 54000   & 250000  & 0.45025 \\
            18 & 80000   & 240000  & 0.46315 \\
            19 & 120000  & 400000  & 0.56551 \\
            20 & 54000   & 220000  & 0.47653 \\
            21 & 59000   & 250000  & 0.51744 \\
            22 & 55000   & 70000   & 0.48923 \\
            23 & 25000   & 45000   & 0.52166 \\
            24 & 350000  & 70000   & 0.47205 \\
            25 & 50000   & 100000  & 0.45293 \\
            %\bottomrule
        \end{tabular}}
    \end{minipage}%
    \hfill
    \begin{minipage}{0.48\textwidth}
        \centering
        \resizebox{\textwidth}{!} {%
        \begin{tabular}{cccc}
            \toprule
            \small{\textbf{Agri-Owner}} & \small{\textbf{Harvest Quantity}} &  \small{\textbf{Processing Capacity}} & \small{\textbf{Processing Cost}} \\
            \midrule
            26 & 80000   & 360000  & 0.45233 \\
            27 & 35000   & 90000   & 0.50965 \\
            28 & 80000   & 100000  & 0.48968 \\
            29 & 73000   & 200000  & 0.48912 \\
            30 & 180000  & 300000  & 0.53758 \\
            31 & 75000   & 100000  & 0.50375 \\
            32 & 81000   & 200000  & 0.50093 \\
            33 & 60000   & 200000  & 0.54263 \\
            34 & 98000   & 600000  & 0.47087 \\
            35 & 400000  & 1000000 & 0.51590 \\
            36 & 100000  & 300000  & 0.52005 \\
            37 & 56000   & 100000  & 0.48769 \\
            38 & 50000   & 200000  & 0.50147 \\
            39 & 100000  & 200000  & 0.46002 \\
            40 & 27000   & 200000  & 0.46056 \\
            41 & 60000   & 350000  & 0.49815 \\
            42 & 28000   & 800000  & 0.51667 \\
            43 & 42000   & 500000  & 0.57256 \\
            44 & 60000   & 300000  & 0.47499 \\
            45 & 35000   & 150000  & 0.50062 \\
            46 & 200000  & 300000  & 0.49253 \\
            47 & 100000  & 500000  & 0.43701 \\
            48 & 80000   & 720000  & 0.48183 \\
            49 & 130000  & 80000   & 0.44116 \\
            %\bottomrule
        \end{tabular}}
    \end{minipage}
    \caption{Harvest quantity, processing capacity, and processing cost}
    \label{tab:data22}
\end{table}

In Table \ref{tab:data22} we show the Harvest Quantity and Processing Capacity in kilograms, and the Processing Cost in euros. In the original database, the processing costs for each agri-owner are not precise but fall within a range [0.44, 0.55]. For this reason, and to apply our model, we simulated these costs using a normal distribution with a mean of 0.495 and standard deviation of 0.03. The normal distribution is widely used in modeling natural and economic phenomena due to its flexibility in representing a wide variety of data distributions. This allows us to capture the natural variability that may exist in processing costs among different companies.

The market price $p$ is found in the link called "the farmer's portal"\footnote{https://agroalbania.al/single-product.php?prod=29} from the Ministry of Agriculture and Rural Development in Albania, which is a digital platform that provides information and services in the agricultural sector. This portal updates the daily prices for fruits and vegetables according to the market in each region. There is a range between the minimum and the maximum prices. In our application we take $p=0.70 $\euro \ for illustration.

Based on the data from Table \ref{tab:data22}, we conduct an initial case study involving 7 Agri-Owners.

\begin{casess}
We select Agri-Owners 14, 22, 23, 24, 25, 31, and 32, along with their respective costs, processing capacities, and harvest quantities, who decide to cooperate (see Table \ref{tab:subset_sorted}). We have chosen this initial example for its visual appeal, as we achieve maximum bounds of positive compensation for both the BTC allocation and the CRC allocation.
\begin{table}[h]
    \centering
    \begin{tabular}{cccc}
        \toprule
        \textbf{Agri-Owner} & \textbf{Harvest Quantity} & \textbf{Processing Capacity} & \textbf{Processing Cost} \\
        \midrule
        14 & 120000  & 155000  & 0.52802 \\
        22 & 55000   & 70000   & 0.48923 \\
        23 & 25000   & 45000   & 0.52166 \\
        24 & 350000  & 70000   & 0.47205 \\
        25 & 50000   & 100000  & 0.45293 \\
        31 & 75000   & 100000  & 0.50375 \\
        32 & 81000   & 200000  & 0.50093 \\
        \bottomrule
    \end{tabular}
    \caption{Subset of data: Harvest quantity, processing capacity, and processing cost}
    \label{tab:subset_sorted}
\end{table}

In this scenario, the profit of the grand coalition would be: $v(N)=182,832$, $\bar{\alpha} =0.03869$ and $\bar{\beta} = 0.147058$. Applying our procedure, we present all the allocations studied In Table \ref{tabla:comp_mod2_with_diff}.
\begin{table}[h!]
    \centering
    \begin{tabular}{|c|c|c|c|c|c|}
        %\hline
        \textbf{Agri-Owner} & $\gamma (v)$ & $T(v,\frac{\bar{\alpha}}{2})$ & $R(v,\frac{\bar{\beta}}{2})$ & $HTR(v,\frac{\bar{\alpha}}{2},\frac{\bar{\beta}}{2})$ & $HTR(v,\frac{\bar{\alpha}}{2},\frac{\bar{\beta}}{2}) - \gamma (v)$ \\ \hline
        14 & 38296 & 36814 & 32664 & 31182 & -5632 \\ \hline
        22 & 17295 & 16626 & 14752 & 14082 & -2544 \\ \hline
        23 & 11118 & 10688 & 9483 & 9053 & -1635 \\ \hline
        24 & 17295 & 16626 & 41639 & 40969 & +24343 \\ \hline
        25 & 24707 & 30825 & 21074 & 27192 & -3633 \\ \hline
        31 & 24707 & 23751 & 21074 & 20118 & -3633 \\ \hline
        32 & 49414 & 47502 & 42147 & 40235 & -7267 \\ \hline
    \end{tabular}
    \caption{Comparison of the CO, BTC, CRC, and HTR allocations with difference}
    \label{tabla:comp_mod2_with_diff}
\end{table}

The farmer who provides lower processing costs is Agri-Owner 25 and he will receive additional compensation for his contributions. Farmers with surplus processing capacity, such as Agri-Owner 14, 22, 23,25, 31 and 32, must compensate those who help process their excess harvest. Agri-Owner 24, who has a high harvest quantity but lower processing capacity, benefits significantly from the cooperation, receiving the highest positive compensation adjustment (+243,43). 

The HTR allocation method ensures that the combined efforts of all farmers lead to higher overall profits, as seen in the positive adjustments for key contributors like Agri-Owner 24. 
\end{casess}
%\medskip

All calculations in this section were performed using Matlab (online version). The following case illustrates the scenario where 22 Agri-Owners decide to cooperate.

\begin{casess}
We selected the Agri-Owner 7,10,11,24 and 49 since they are the only ones with more harvest than processing capacity; the rest will be chosen randomly. We only consider 22 agri-owners from the original database of 49 due to limitations in Matlab Online, which has restricted computational capacity for online users. Table \ref{tab:selected_data2} shows the selected Agri-Owners.

\begin{table}[h!]
    \centering
    \resizebox{0.6\textwidth}{!} {%
    \begin{tabular}{cccc}
        \toprule
        \small{\textbf{Agri-Owner}} & \small{\textbf{Harvest Quantity}} & \small{\textbf{Processing Capacity}} & \small{\textbf{Processing Cost}} \\
        \midrule
        3  & 96000   & 300000  & 0.45858 \\
        5  & 63000   & 120000  & 0.49479 \\
        7  & 80000   & 60000   & 0.47191 \\
        9  & 60000   & 70000   & 0.48823 \\
        10 & 100000  & 87000   & 0.52852 \\
        11 & 120000  & 70000   & 0.46233 \\
        13 & 75000   & 500000  & 0.51158 \\
        14 & 120000  & 155000  & 0.52802 \\
        15 & 72000   & 200000  & 0.54133 \\
        20 & 54000   & 220000  & 0.47653 \\
        24 & 350000  & 70000   & 0.47205 \\
        25 & 50000   & 100000  & 0.45293 \\
        26 & 80000   & 360000  & 0.45233 \\
        27 & 35000   & 90000   & 0.50965 \\
        28 & 80000   & 100000  & 0.48968 \\
        35 & 400000  & 1000000 & 0.51590 \\
        36 & 100000  & 300000  & 0.52005 \\
        38 & 50000   & 200000  & 0.50147 \\
        40 & 27000   & 200000  & 0.46056 \\
        45 & 35000   & 150000  & 0.50062 \\
        48 & 80000   & 720000  & 0.48183 \\
        49 & 130000  & 80000   & 0.44116 \\
        \bottomrule
    \end{tabular}%
    }
    \caption{Harvest quantity, processing capacity, and processing cost for selected Agri-Owners}
    \label{tab:selected_data2}
\end{table}

In this case, the maximum profit generated by the cooperative for all agri-owners is $v(N)=584202$, $\bar{\alpha} =0.02157$ and $\bar{\beta} =0$. Applying the procedure described in the previous section, we will stick with the BTC allocation since $R(v,\frac{\bar{\beta}}{2})=\gamma (v)$. We obtain Table \ref{tabla:comp_mod4} where the allocations CO and BTC can be found. 
%\begin{table}[h!]
%    \centering
%    \begin{tabular}{|c|c|c|c|c|}
%        %\hline
%        \textbf{Agri-Owner} & $\gamma (v)$ & $T(v,\frac{\bar{\alpha}}{2})$ & $R(v,\frac{\bar{\beta}}{2})$ & $HTR(v,\frac{\bar{\alpha}}{2},\frac{\bar{\beta}}{2})$ \\ \hline
%        3  & 24849 & 24312 & 24849 & 24312 \\ \hline
%        5  & 16307 & 15955 & 16307 & 15955 \\ \hline
%        7  & 20707 & 20260 & 20707 & 20260 \\ \hline
%        9  & 15530 & 15195 & 15530 & 15195 \\ \hline
%        10 & 25884 & 25326 & 25884 & 25326 \\ \hline
%        11 & 31061 & 30391 & 31061 & 30391 \\ \hline
%        13 & 19413 & 18994 & 19413 & 18994 \\ \hline
%        14 & 31061 & 30391 & 31061 & 30391 \\ \hline
%        15 & 18636 & 18234 & 18636 & 18234 \\ \hline
%        20 & 13977 & 13676 & 13977 & 13676 \\ \hline
%        24 & 90594 & 88639 & 90594 & 88639 \\ \hline
%        25 & 12942 & 12663 & 12942 & 12663 \\ \hline
%        26 & 20707 & 20260 & 20707 & 20260 \\ \hline
%        27 & 9059  & 8864  & 9059  & 8864  \\ \hline
%        28 & 20707 & 20260 & 20707 & 20260 \\ \hline
%        35 & 103536 & 101302 & 103536 & 101302 \\ \hline
%        36 & 25884 & 25326 & 25884 & 25326 \\ \hline
%        38 & 12942 & 12663 & 12942 & 12663 \\ \hline
%        40 & 6989  & 6838  & 6989  & 6838  \\ \hline
%        45 & 9059  & 8864  & 9059  & 8864  \\ \hline
%        48 & 20707 & 20260 & 20707 & 20260 \\ \hline
%        49 & 33649 & 45528 & 33649 & 45528 \\ \hline
%    \end{tabular}
%    \caption{Comparison of the CO, BTC, CRC, and HTR allocations (sorted by Agri-Owner)}
%    \label{tabla:comp_mod3}
%\end{table}

\begin{table}[h!]
    \centering
    \begin{tabular}{|c|c|c|c|}
        %\hline
        \textbf{Agri-Owner} & $\gamma (v)$ & $T(v,\frac{\bar{\alpha}}{2})$ & $T(v,\frac{\bar{\alpha}}{2}) - \gamma(v)$ \\ \hline
        3  & 24849 & 24312 & $-537$ \\ \hline
        5  & 16307 & 15955 & $-352$ \\ \hline
        7  & 20707 & 20260 & $-447$ \\ \hline
        9  & 15530 & 15195 & $-335$ \\ \hline
        10 & 25884 & 25326 & $-558$ \\ \hline
        11 & 31061 & 30391 & $-670$ \\ \hline
        13 & 19413 & 18994 & $-419$ \\ \hline
        14 & 31061 & 30391 & $-670$ \\ \hline
        15 & 18636 & 18234 & $-402$ \\ \hline
        20 & 13977 & 13676 & $-301$ \\ \hline
        24 & 90594 & 88639 & $-1955$ \\ \hline
        25 & 12942 & 12663 & $-279$ \\ \hline
        26 & 20707 & 20260 & $-447$ \\ \hline
        27 & 9059  & 8864  & $-195$ \\ \hline
        28 & 20707 & 20260 & $-447$ \\ \hline
        35 & 103536 & 101302 & $-2234$ \\ \hline
        36 & 25884 & 25326 & $-558$ \\ \hline
        38 & 12942 & 12663 & $-279$ \\ \hline
        40 & 6989  & 6838  & $-151$ \\ \hline
        45 & 9059  & 8864  & $-195$ \\ \hline
        48 & 20707 & 20260 & $-447$ \\ \hline
        49 & 33649 & 45528 & $+11879$ \\ \hline
    \end{tabular}
    \caption{Comparison of the CO and BTC allocations}
    \label{tabla:comp_mod4}
\end{table}

$\bar{\alpha} =0.02157$: This parameter represents the maximum percentage rate used to tax farmers who do not have the lowest processing costs. It is a measure of the contribution to cost savings that these farmers need to pay to those with the best technology. 

 $\bar{\beta} =0$: the Crop Reward Compensation allocation (CRC) does not apply in this scenario, meaning the BTC allocation is the sole focus. 

Agri-Owner 49, who has a high harvest quantity but lower processing capacity, benefits significantly from the cooperative effort, receiving substantial compensation (+11,879) for sharing their excess harvest and benefiting from others' processing capacities. 

Agri-Owner 35 and Agri-Owner 24, with large capacities, contribute more towards compensating those with efficient technology, as seen in their negative differences in the BTC allocation (-2234 and -1955 respectively). 

\end{casess}

\section{\protect\bigskip Conclusions and further research}
This paper introduces an innovative cooperative framework for agricultural crop processing using cooperative game theory. The proposed model highlights how collaboration among firms can effectively address the challenges encountered in competitive and resource-limited environments. By incorporating key concepts such as transferable utility games and core-stable profit allocations, the framework ensures equitable compensation for all participating firms, reflecting their contributions to cost savings and enhanced processing outcomes.

The findings highlight the effectiveness of cooperative strategies, especially in scenarios where mismatches arise between harvesting and processing capacities. The introduction of the HarvestTech (HT) situation enables firms to pool resources, technologies, and processing capabilities, optimizing overall profitability while minimizing inefficiencies. This cooperative model goes beyond economic objectives, fostering long-term partnerships that enhance sustainability in agricultural operations.

The Collaborative allocation (CO) ensures proportional compensation based on each firm’s processing capacity or harvest contribution, while the Best Technology Compensation (BTC) and Crop Reward Compensation (CRC) allocations address additional complexities such as technology sharing and capacity mismatches. These allocation mechanisms foster fairness and stability within the coalition, incentivizing active participation and resource sharing among firms for mutual gain.

The application of the HarvestTech (HT) model in the Korça apple farming case studies demonstrates the practical value of cooperative strategies in agricultural processing. By enabling farmers to share processing technologies and capacities, the HT model minimizes operational costs, prevents surplus harvest losses, and ensures fair profit distribution. Through a procedure designed based on the CO, BTC, and CRC allocations, participants with advanced technologies or surplus harvests receive appropriate compensation, fostering collaboration. The results show that cooperative efforts increase overall profitability while maintaining stable allocations, ensuring that no participant has an incentive to leave the grand coalition. 

In conclusion, this study highlights the importance of cooperative game theory as a valuable tool for addressing the challenges faced by agricultural firms, such as fluctuating harvests and capacity mismatches. The study demonstrates that fair and stable profit allocations encourage long-term collaboration, promoting sustainable agricultural practices. The HT framework offers a solid foundation for optimizing agricultural supply chains, ensuring profitability, sustainability, and resilience in dynamic markets. Moreover, the findings show that equitable profit sharing reinforces firms' willingness to cooperate, fostering a stable and collaborative environment.

In this context, we propose several future lines of research to further explore and expand upon our innovative approach to agricultural crop processing through inter-firm cooperation. One area of investigation involves the application and development of Biform HarvestTech (HT) games within the agricultural sector, examining their use in modeling more complex strategic interactions among firms. Additionally, analyzing scenarios where each firm decides whether to cooperate and demonstrating that the profile in which all firms choose to cooperate constitutes a Nash equilibrium in the first stage, given the use of the HTR rule for allocation in the second stage, is crucial. Developing more comprehensive models that account for a wider range of costs and variables will enhance the applicability of the model to more complex and realistic agricultural processing situations. Furthermore, studying models where prices and costs are functions will allow for a more dynamic and realistic representation of market and cost structures. Lastly, investigating models where the quantities of harvest and production capacities are not deterministic, including exploring stochastic models and fuzzy logic approaches, will better handle uncertainty and variability in agricultural production. These research directions aim to refine and expand the current model, increasing its applicability and robustness in real-world agricultural processing scenarios.

\subsection*{Acknowledgements}
This work is part of the R+D+I project grant PID2022-137211NB-100 that was funded by MCIN/AEI/10.13039/50110001133/ and by "ERDF A a way of making EUROPE"/UE. This research was also funded by project PROMETEO/2021/063 from the Comunidad Valenciana. 

\subsection*{Declarations}
\textbf{Conflict of interest} Not applicable.

\bibliography{bibliography}

\begin{thebibliography}{}

\bibitem[Bernstein et~al., 2015]{bernstein2015cooperation}
Bernstein, F., K{\"o}k, A.~G., and Meca, A. (2015).
\newblock Cooperation in assembly systems: The role of knowledge sharing networks.
\newblock {\em European Journal of Operational Research}, 240(1):160--171.

\bibitem[Bondareva, 1963]{bondareva1963some}
Bondareva, O.~N. (1963).
\newblock Some applications of linear programming methods to the theory of cooperative games.
\newblock {\em Problemy Kibernet}, 10:119.

\bibitem[Canaj and Mehmeti, 2024]{canaj2024energy}
Canaj, K. and Mehmeti, A. (2024).
\newblock Energy consumption and environmental impacts in western balkan apple production: A case study of the kor{\c{c}}a province, albania.
\newblock {\em Applied Fruit Science}, 66(2):417--429.

\bibitem[Cruijssen et~al., 2007]{cruijssen2007horizontal}
Cruijssen, F., Dullaert, W., and Fleuren, H. (2007).
\newblock Horizontal cooperation in transport and logistics: a literature review.
\newblock {\em Transportation journal}, 46(3):22--39.

\bibitem[Curiel et~al., 2002]{curiel2002sequencing}
Curiel, I., Hamers, H., and Klijn, F. (2002).
\newblock Sequencing games: a survey.
\newblock In {\em Chapters in game theory: in honor of Stef Tijs}, pages 27--50. Springer.

\bibitem[Fiestras-Janeiro et~al., 2012]{fiestras2012cost}
Fiestras-Janeiro, M.~G., Garc{\'\i}a-Jurado, I., Meca, A., and Mosquera, M. (2012).
\newblock Cost allocation in inventory transportation systems.
\newblock {\em Top}, 20:397--410.

\bibitem[Fiestras-Janeiro et~al., 2011]{fiestras2011cooperative}
Fiestras-Janeiro, M.~G., Garc{\'\i}a-Jurado, I., Meca, A., and Mosquera, M.~A. (2011).
\newblock Cooperative game theory and inventory management.
\newblock {\em European Journal of Operational Research}, 210(3):459--466.

\bibitem[Fiestras-Janeiro et~al., 2013]{fiestras2013new}
Fiestras-Janeiro, M.~G., Garc{\'\i}a-Jurado, I., Meca, A., and Mosquera, M.~A. (2013).
\newblock A new cost allocation rule for inventory transportation systems.
\newblock {\em Operations Research Letters}, 41(5):449--453.

\bibitem[Fiestras-Janeiro et~al., 2014]{fiestras2014centralized}
Fiestras-Janeiro, M.~G., Garc{\'\i}a-Jurado, I., Meca, A., and Mosquera, M.~A. (2014).
\newblock Centralized inventory in a farming community.
\newblock {\em Journal of Business Economics}, 84:983--997.

\bibitem[Fiestras-Janeiro et~al., 2015]{fiestras2015cooperation}
Fiestras-Janeiro, M.~G., Garc{\'\i}a-Jurado, I., Meca, A., and Mosquera, M.~A. (2015).
\newblock Cooperation on capacitated inventory situations with fixed holding costs.
\newblock {\em European Journal of Operational Research}, 241(3):719--726.

\bibitem[Guardiola et~al., 2023]{GUARDIOLA2023102889}
Guardiola, L.~A., Meca, A., and Puerto, J. (2023).
\newblock Allocating the surplus induced by cooperation in distribution chains with multiple suppliers and retailers.
\newblock {\em Journal of Mathematical Economics}, 108:102889.

\bibitem[Guardiola et~al., 2007]{guardiola2007cooperation}
Guardiola, L.~A., Meca, A., and Timmer, J. (2007).
\newblock Cooperation and profit allocation in distribution chains.
\newblock {\em Decision support systems}, 44(1):17--27.

\bibitem[Hezarkhani et~al., 2021]{hezarkhani2021collaboration}
Hezarkhani, B., Slikker, M., and Woensel, T.~V. (2021).
\newblock Collaboration in transport and logistics networks.
\newblock {\em Network Design with Applications to Transportation and Logistics}, pages 627--662.

\bibitem[J{\o}rgensen and Zaccour, 2014]{jorgensen2014survey}
J{\o}rgensen, S. and Zaccour, G. (2014).
\newblock A survey of game-theoretic models of cooperative advertising.
\newblock {\em European journal of operational Research}, 237(1):1--14.

\bibitem[Meca and Timmer, 2008]{meca2008supply}
Meca, A. and Timmer, J. (2008).
\newblock Supply chain collaboration.
\newblock {\em Kordic, V.(Ed.), Supply Chain, Theory and Applications}, pages 1--18.

\bibitem[Nagarajan and So{\v{s}}i{\'c}, 2008]{nagarajan2008game}
Nagarajan, M. and So{\v{s}}i{\'c}, G. (2008).
\newblock Game-theoretic analysis of cooperation among supply chain agents: Review and extensions.
\newblock {\em European journal of operational research}, 187(3):719--745.

\bibitem[Osmani and Kambo, 1999]{osmani1999efficiency}
Osmani, M. and Kambo, A. (1999).
\newblock Efficiency of apple small-scale farming in albania-a stochastic frontier approach.
\newblock {\em New Medit}, 18(2):71--88.

\bibitem[Osmani and Kambo, 2019]{osmani2019small}
Osmani, M. and Kambo, A. (2019).
\newblock Small-scale apple farmers'willingness to invest-the case of kor{\c{c}}a region farmers in albania.
\newblock {\em Acta Universitatis Agriculturae et Silviculturae Mendelianae Brunensis}, 67(1).

\bibitem[Rzeczycki, 2022]{rzeczycki2022supply}
Rzeczycki, A. (2022).
\newblock Supply chain decision making with use of game theory.
\newblock {\em Procedia Computer Science}, 207:3988--3997.

\bibitem[Shapley, 1971]{shapley1971cores}
Shapley, L.~S. (1971).
\newblock Cores of convex games.
\newblock {\em International journal of game theory}, 1:11--26.

\bibitem[Shapley et~al., 1967]{shapley1967balanced}
Shapley, L.~S. et~al. (1967).
\newblock On balanced sets and cores.
\newblock {\em Naval Research Logistics Quarterly}, 14(4):453--460.

\bibitem[Shapley and Shubik, 1969]{shapley1969market}
Shapley, L.~S. and Shubik, M. (1969).
\newblock On market games.
\newblock {\em Journal of Economic Theory}, 1(1):9--25.

\bibitem[Shkreli and Imami, 2019]{shkreli2019citrus}
Shkreli, E. and Imami, D. (2019).
\newblock Citrus sector study.
\newblock {\em ASF Project Office: Tiran{\"e}, Albania}.

\bibitem[Tavanayi et~al., 2021]{tavanayi2021cooperative}
Tavanayi, M., Hafezalkotob, A., and Valizadeh, J. (2021).
\newblock Cooperative cellular manufacturing system: A cooperative game theory approach.
\newblock {\em Scientia Iranica}, 28(5):2769--2788.

\bibitem[Zheng et~al., 2022]{zheng2022cooperative}
Zheng, X.-X., Guo, J., Jia, F., and Zhang, S. (2022).
\newblock Cooperative game theory approach to develop an incentive mechanism for biopesticide adoption through farmer producer organizations.
\newblock {\em Journal of Environmental Management}, 319:115696.

\end{thebibliography}
\bibliographystyle{apalike}

 \newpage
\section*{Appendix - Proofs}

\begin{proof}[Proof of Proposition \protect\ref{prop1}]
(i) Take $S \subseteq N$. Then, $v(S) = (p - c_{S}) \cdot \min\{K_{S}, Q_{S}\} \geq 0$ since $p - c_{S} > 0$ and $K_{S}, Q_{S} \geq 0$.

(ii) Take $S, T \subseteq N$ such that $S \cap T = \emptyset$. We know from the definition of the model that $K_{S}, Q_{S}, K_{T}, Q_{T} \geq 0$. First, we will prove that

\begin{equation}\label{eq0}
\min\{K_{S \cup T}, Q_{S \cup T}\} \geq \min\{K_{S}, Q_{S}\} + \min\{K_{T}, Q_{T}\}.
\end{equation}

To do this, we will consider the following cases:

\begin{itemize}
    \item \textbf{Case 1:} $\min\{K_{S \cup T}, Q_{S \cup T}\} = K_{S \cup T}$.
    \begin{itemize}
        \item 1.1 $K_{S} \leq Q_{S}$ and $K_{T} \leq Q_{T}$. Then, (\ref{eq0}) is satisfied with equality.
        \item 1.2 $K_{S} > Q_{S}$ and $K_{T} \leq Q_{T}$. Then, $K_{S \cup T} = K_{S} + K_{T} > Q_{S} + K_{T}$ satisfying (\ref{eq0}).
        \item 1.3 $K_{S} \leq Q_{S}$ and $K_{T} > Q_{T}$. Then, $K_{S \cup T} = K_{S} + K_{T} > K_{S} + Q_{T}$ satisfying (\ref{eq0}).
        \item 1.4 $K_{S} > Q_{S}$ and $K_{T} > Q_{T}$. This case is not possible since $K_{S \cup T} \leq Q_{S \cup T}$.
    \end{itemize}
    \item \textbf{Case 2:} $\min\{K_{S \cup T}, Q_{S \cup T}\} = Q_{S \cup T}$. This is proved analogously to the previous case.
\end{itemize}

Therefore, (\ref{eq0}) is satisfied. Additionally, $p - c_{S \cup T} \geq p - c_{S} > 0$ and $p - c_{S \cup T} \geq p - c_{T} > 0$ since $c_{S \cup T} = \min\{c_{i}\}$. Hence, the following inequality is satisfied:
$$(p - c_{S \cup T}) \cdot \min\{K_{S \cup T}, Q_{S \cup T}\} \geq (p - c_{S}) \cdot \min\{K_{S}, Q_{S}\} + (p - c_{T}) \cdot \min\{K_{T}, Q_{T}\};$$

The above inequality is equivalent to: $v(S \cup T) \geq v(S) + v(T)$. We conclude that the game is superadditive.

Finally, (iii) follows from (i) and (ii).
\end{proof}

\bigskip

\begin{proof}[Proof of Theorem \protect\ref{prop2}]
Suppose that $K_{N} \leq Q_{N}$ (The proof for the case $K_{N} > Q_{N}$ is analogous). Then, the CO allocation is efficient since $\sum_{i \in N} \gamma_{i}(v) = (p - c_{N}) \cdot K_{N} = v(N)$.

Now, we show that this allocation satisfies coalition rationality. For this, consider a coalition $S \subseteq N$. Then, $\sum_{i \in S} \gamma_{i}(v) = (p - c_{N}) \cdot K_{S} \geq (p - c_{S}) \cdot K_{S} \geq (p - c_{S}) \cdot \min\{K_{S}, Q_{S}\} = v(S)$.

Therefore, we can conclude that $\gamma(v) \in \text{Core}(N, v)$.
\end{proof}

\bigskip

\begin{proof}[Proof of Theorem \protect\ref{theo1}]
It is evident that this distribution is efficient. We check coalition
stability for each type of coalition: If $S\subseteq M$ then, $$\sum_{i\in
S}T_{i}(v,\alpha )=\sum_{i\in S}\gamma _{i}(v)+\left\vert S\right\vert \frac{%
\alpha }{\left\vert M\right\vert }\sum_{j\in N/M}\gamma _{j}(v)\geq
\sum_{i\in S}\gamma _{i}(v)\geq v(S)$$ since $\gamma (v)\in Core(N,v).$ If $%
S\subseteq N\backslash M$ then it should be satisfied that: 
\begin{eqnarray*}
\sum_{i\in S}T_{i}(v,\alpha ) &=&\sum_{i\in S}\gamma _{i}(v)-\alpha
\sum_{i\in S}\gamma _{i}(v)\geq v(S); \\
\alpha \sum_{i\in S}\gamma _{i}(v) &\leq &\sum_{i\in S}\gamma _{i}(v)-v(S);
\\
\alpha  &\leq &1-\frac{v(S)}{\sum_{i\in S}\gamma _{i}(v)}
\end{eqnarray*}%
Hence, $\alpha \leq $ $\underset{S\subseteq N\backslash M}{\min }\left\{ 1-%
\frac{v(S)}{\sum_{i\in S}\gamma _{i}(v)}\right\} .$ Finally, if $S\cap M\neq
\emptyset $ and $S\cap (N\backslash M)\neq \emptyset $ then it should be
satisfied that: $\sum_{i\in S}T_{i}(v,\alpha )\geq v(S)$, i.e.:%
\begin{eqnarray}\label{eq1}
\sum_{i\in S\cap (N\backslash M)}\gamma _{i}(v)-\alpha \sum_{i\in S\cap
(N\backslash M)}\gamma _{i}(v)+\sum_{i\in S\cap M}\gamma _{i}(v)+\left\vert
S\cap M\right\vert \frac{\alpha }{\left\vert M\right\vert }\sum_{j\in
N/M}\gamma _{j}(v) &\geq &v(S); \notag\\
\sum_{i\in S}\gamma _{i}(v)+\alpha \left( \frac{\left\vert S\cap
M\right\vert }{\left\vert M\right\vert }\sum_{j\in N/M}\gamma
_{j}(v)-\sum_{i\in S\cap (N\backslash M)}\gamma _{i}(v)\right)  &\geq &v(S);
\end{eqnarray}

if $\frac{\left\vert S\cap M\right\vert }{\left\vert M\right\vert }%
\sum_{j\in N/M}\gamma _{j}(v)-\sum_{i\in S\cap (N\backslash M)}\gamma
_{i}(v)\geq 0$ then the above inequality is satisfied. Therefore, we
consider the set 
\begin{equation*}
\Lambda :=\left\{ S\subseteq N:S\cap M\neq \emptyset ,S\cap (N\backslash
M)\neq \emptyset \text{ and}\sum_{i\in S\cap (N\backslash M)}\gamma _{i}(v)>%
\frac{\left\vert S\cap M\right\vert }{\left\vert M\right\vert }\sum_{j\in
N/M}\gamma _{j}(v)\right\}. 
\end{equation*}

Then, (\ref{eq1}) it is equivalent to:%
\begin{eqnarray*}
\alpha \left( \sum_{i\in S\cap (N\backslash M)}\gamma _{i}(v)-\frac{%
\left\vert S\cap M\right\vert }{\left\vert M\right\vert }\sum_{j\in
N/M}\gamma _{j}(v)\right)  &\leq &\sum_{i\in S}\gamma _{i}(v)-v(S); \\
\alpha  &\leq &\frac{\sum_{i\in S}\gamma _{i}(v)-v(S)}{\sum_{i\in S\cap
(N\backslash M)}\gamma _{i}(v)-\frac{\left\vert S\cap M\right\vert }{%
\left\vert M\right\vert }\sum_{j\in N/M}\gamma _{j}(v)}.
\end{eqnarray*}%
Hence, $\alpha \leq $ $\underset{S\in \Lambda }{\min }\left\{ \frac{%
\sum_{i\in S}\gamma _{i}(v)-v(S)}{\sum_{i\in S\cap (N\backslash M)}\gamma
_{i}(v)-\frac{\left\vert S\cap M\right\vert }{\left\vert M\right\vert }%
\sum_{j\in N/M}\gamma _{j}(v)}\right\}$, and since all partial results have
been obtained through equivalences, we can conclude by ensuring that the BTC
allocation belongs to the core if and only if: 
\begin{equation*}
0\leq \alpha \leq \min \left\{ \underset{S\subseteq N\backslash M}{\min }%
\left\{ 1-\frac{v(S)}{\underset{i\in S}{\sum }\gamma _{i}(v)}\right\} ,%
\underset{S\in \Lambda }{\min }\left\{ \frac{\sum_{i\in S}\gamma _{i}(v)-v(S)%
}{\underset{i\in S\cap (N\backslash M)}{\sum }\gamma _{i}(v)-\frac{%
\left\vert S\cap M\right\vert }{\left\vert M\right\vert }\underset{j\in N/M}{%
\sum }\gamma _{j}(v)}\right\} \right\} 
\end{equation*}
\end{proof}
\bigskip

\begin{proof}[Proof of Theorem \protect\ref{theo2}]
It is evident that this distribution is efficient. Now, we check coalition
stability for each type of coalition. If $S\subseteq E$ it is trivial. If $%
S\subseteq H$ then: 
\begin{equation*}
\sum_{i\in S}R_{i}(v,\alpha )=\sum_{i\in S}\gamma _{i}(v)+\beta \frac{%
Q_{S}-K_{S}}{Q_{H}-K_{H}}\sum_{j\in N\backslash H}\gamma _{j}(v)\geq
\sum_{i\in S}\gamma _{i}(v)\geq v(S),
\end{equation*}
since $\frac{Q_{S}-K_{S}}{Q_{H}-K_{H}}$ always is positive and $\gamma
(v)\in Core(N,v).$

If $S\subseteq N\backslash \left( H\cup E\right) $ then it should be
satisfied that: 
\begin{eqnarray*}
\sum_{i\in S}R_{i}(v,\alpha ) &=&\sum_{i\in S}\gamma _{i}(v)-\beta
\sum_{i\in S}\gamma _{i}(v)\geq v(S); \\
\beta \sum_{i\in S}\gamma _{i}(v) &\leq &\sum_{i\in S}\gamma _{i}(v)-v(S); \\
\beta  &\leq &1-\frac{v(S)}{\sum_{i\in S}\gamma _{i}(v)}.
\end{eqnarray*}

Hence, 
\begin{equation}\label{eq2}
\beta \leq \underset{S\subseteq N\backslash \left( H\cup E\right) }{\min }%
\left\{ 1-\frac{v(S)}{\sum_{i\in S}\gamma _{i}(v)}\right\} 
\end{equation}

If $S\subseteq H\cup E$ then $\sum_{i\in S}R_{i}(v,\alpha )=\sum_{i\in
S}\gamma _{i}(v)+\frac{Q_{S\cap H}-K_{S\cap H}}{Q_{H}-K_{H}}\beta \sum_{j\in
N\backslash H}\gamma _{j}(v)\geq \sum_{i\in S}\gamma _{i}(v)\geq v(S).$

If $S\subseteq N\backslash H$ with $S\backslash E\neq \emptyset $ (this implies that $S\cap \left( N\backslash \left( H\cup E\right) \right) \neq
\emptyset $) then it should be satisfied that: 
\begin{eqnarray*}
\sum_{i\in S}R_{i}(v,\alpha ) &=&\sum_{i\in S}\gamma _{i}(v)-\beta
\sum_{i\in S\cap \left( N\backslash \left( H\cup E\right) \right) }\gamma
_{i}(v)\geq v(S); \\
\beta \sum_{i\in S\cap \left( N\backslash \left( H\cup E\right) \right)
}\gamma _{i}(v) &\leq &\sum_{i\in S}\gamma _{i}(v)-v(S); \\
\beta  &\leq &1-\frac{v(S)}{\sum_{i\in S\cap \left( N\backslash \left( H\cup
E\right) \right) }\gamma _{i}(v)}
\end{eqnarray*}%
Hence, 
\begin{equation}\label{eq3}
\beta \leq \underset{S\subseteq N\backslash H:S\backslash E\neq \emptyset }{%
\min }\left\{ 1-\frac{v(S)}{\sum_{i\in S\cap \left( N\backslash \left( H\cup
E\right) \right) }\gamma _{i}(v)}\right\} .
\end{equation}
Note that (\ref{eq2}) and (\ref{eq3}) are equivalent to: $\beta \leq \underset{S\subseteq N\backslash
H:S\backslash E\neq \emptyset }{\min }\left\{ 1-\frac{v(S)}{\sum_{i\in
S\backslash E}\gamma _{i}(v)}\right\} .$ 

Finally, if $S\cap \left( N\backslash \left( H\cup E\right) \right) \neq
\emptyset $ and $S\cap H\neq \emptyset $ then it should be satisfied that: $%
\sum_{i\in S}R_{i}(v,\alpha )\geq v(S)$, i.e.:%

\begin{eqnarray*}
&&\sum_{i\in S\cap \left( N\backslash \left( H\cup E\right) \right) }\gamma
_{i}(v)+\sum_{i\in S\cap E}\gamma _{i}(v)-\beta \sum_{i\in S\cap \left(
N\backslash \left( H\cup E\right) \right) }\gamma _{i}(v)+  \notag \\&&
\sum_{i\in S\cap H}\gamma _{i}(v)+\beta \frac{Q_{S\cap H}-K_{S\cap H}}{%
Q_{H}-K_{H}}\sum_{j\in N\backslash \left( H\cup E\right) }\gamma _{j}(v)
\geq v(S);
\end{eqnarray*}%
\begin{equation}\label{eq4}
\sum_{i\in S}\gamma _{i}(v)+\beta \left( \frac{Q_{S\cap H}-K_{S\cap H}}{%
Q_{H}-K_{H}}\sum_{j\in N\backslash \left( H\cup E\right) }\gamma
_{j}(v)-\sum_{i\in S\cap \left( N\backslash \left( H\cup E\right) \right)
}\gamma _{i}(v)\right)  \geq v(S); 
\end{equation}

if $\frac{Q_{S\cap H}-K_{S\cap H}}{Q_{H}-K_{H}}\sum_{j\in N\backslash \left(
H\cup E\right) }\gamma _{j}(v)-\sum_{i\in S\cap \left( N\backslash \left(
H\cup E\right) \right) }\gamma _{i}(v)\geq 0$ then the above inequality is
satisfied. Therefore, we consider the set 

\begin{equation*}
\Pi :=\left\{ S\subseteq N:%
\begin{array}{c}
S\cap \left( N\backslash \left( H\cup E\right) \right) \neq \emptyset ,S\cap
H\neq \emptyset \text{ and} \\ 
\sum_{i\in S\cap \left( N\backslash \left( H\cup E\right) \right) }\gamma
_{i}(v)>\frac{Q_{S\cap H}-K_{S\cap H}}{Q_{H}-K_{H}}\sum_{j\in N\backslash
\left( H\cup E\right) }\gamma _{j}(v)%
\end{array}%
\right\} 
\end{equation*}

Then (\ref{eq4}) it is equivalent to:%
\begin{equation*}
\beta \left( \sum_{i\in S\cap \left( N\backslash \left( H\cup E\right)
\right) }\gamma _{i}(v)-\frac{Q_{S\cap H}-K_{S\cap H}}{Q_{H}-K_{H}}%
\sum_{j\in N\backslash \left( H\cup E\right) }\gamma _{j}(v)\right) \leq
\sum_{i\in S}\gamma _{i}(v)-v(S);
\end{equation*}%
\begin{equation*}
\beta \leq \frac{\sum_{i\in S}\gamma _{i}(v)-v(S)}{\sum_{i\in S\cap \left(
N\backslash \left( H\cup E\right) \right) }\gamma _{i}(v)-\frac{Q_{S\cap
H}-K_{S\cap H}}{Q_{H}-K_{H}}\sum_{j\in N\backslash \left( H\cup E\right)
}\gamma _{j}(v)};
\end{equation*}%
Hence, $\beta \leq $ $\underset{S\in \Pi }{\min }\left\{ \frac{\sum_{i\in
S}\gamma _{i}(v)-v(S)}{\sum_{i\in S\cap \left( N\backslash \left( H\cup
E\right) \right) }\gamma _{i}(v)-\frac{Q_{S\cap H}-K_{S\cap H}}{Q_{H}-K_{H}}%
\sum_{j\in N\backslash \left( H\cup E\right) }\gamma _{j}(v)}\right\} $ and
since all partial results have been obtained through equivalences, we can
conclude by ensuring that the BTC allocation belongs to the core if and only
if: 
\[
\resizebox{1\textwidth}{!}{$
0\leq \beta \leq \min \left\{ \underset{S\subseteq N\backslash H:S\backslash
E\neq \emptyset }{\min }\left\{ 1-\frac{v(S)}{\underset{i\in S\backslash
E}\sum\gamma _{i}(v)}\right\} ,\underset{S\in \Pi }{\min }\left\{ \frac{%
\underset{i\in S}\sum\gamma _{i}(v)-v(S)}{\underset{i\in S\cap \left( N\backslash \left(
H\cup E\right) \right)}\sum\gamma _{i}(v)-\frac{Q_{S\cap H}-K_{S\cap H}}{%
Q_{H}-K_{H}}\underset{j\in N\backslash \left( H\cup E\right)}\sum\gamma _{j}(v)}%
\right\} \right\}
$}
\]
\end{proof}

\bigskip

\begin{proof}[Proof of Theorem \protect\ref{theo3}]
It is evident that if either of the thresholds $\overline{\alpha }$ or $\overline{\beta}$ is zero, the HTR allocation belongs to the core. Therefore, let us assume that both thresholds are greater than zero. Take $\alpha ^{\ast }\leq \frac{\overline{\alpha }}{2}$ and $\beta ^{\ast
}\leq \frac{\overline{\beta }}{2}.$ It is straightforward that HTR allocation is
efficient. Additionally, we can rewrite it as follows:  
\begin{eqnarray*}
HTR(v,\alpha ^{\ast },\beta ^{\ast }) &=&T(v,\alpha ^{\ast })+R(v,\beta
^{\ast })-\gamma (v)=\left( T(v,\alpha ^{\ast })-\frac{1}{2}\gamma
(v)\right) +\left( R(v,\beta ^{\ast })-\frac{1}{2}\gamma (v)\right)  \\
&=&\frac{1}{2}\left( 2T(v,\alpha ^{\ast })-\gamma (v)\right) +\frac{1}{2}%
\left( 2R(v,\beta ^{\ast })-\gamma (v)\right) .
\end{eqnarray*}%
Hence, if we prove that $2T(v,\alpha ^{\ast })-\gamma (v)$ and $2R(v,\beta
^{\ast })-\gamma (v)$ belongs to the core, then their convex combination will also be in the
core. Indeed, it is easy to see that $2T(v,\alpha ^{\ast })-\gamma
(v)=T(v,2\alpha ^{\ast })\in Core(N,v)$ and $2R(v,\beta ^{\ast })-\gamma
(v)=R(v,2\beta ^{\ast })\in Core(N,v)$ since $2\alpha ^{\ast }\leq \overline{%
\alpha }$ and $2\beta ^{\ast }\leq $ $\overline{\beta }$, respectively.
\end{proof}

\end{document}